\documentclass[%
 reprint,
superscriptaddress,
 amsmath,amssymb,
 aps,
]{revtex4-2}

\usepackage{soul}
\usepackage{tikz}
\usetikzlibrary{shapes.geometric, arrows}
\usepackage{graphicx}% Include figure files
\usepackage{dcolumn}% Align table columns on decimal point
\usepackage{bm}% bold math
\usepackage{hyperref}% add hypertext capabilities
\usepackage{url}

\usepackage[version=4]{mhchem}
\usepackage{siunitx}
\usepackage{comment}
\usepackage{tabularx}
\usepackage{booktabs}
\usepackage[multiple]{footmisc}

\usepackage{csquotes}

\begin{document}

\makeatletter
\long\def\@makecaption#1#2{%
  \small
  \vskip\abovecaptionskip
  \leftskip=0pt \rightskip=0pt \parfillskip=0pt plus 1fil
  \textbf{#1.}\ #2\par
  \vskip\belowcaptionskip}
\makeatother

\title{Hamiltonian learning reveals optoelectronic mechanisms across thermodynamic state space in soft semiconductors}%

\author{Frederik Vonhoff}
\affiliation{%
 Physics Department, TUM School of Natural Sciences, Technical University of Munich, 85748 Garching, Germany
 }
 
\author{Jesper R. Pedersen}
\affiliation{%
 Department of Energy Conversion and Storage, Technical University of Denmark, DK-2800 Kgs.~Lyngby, Denmark
 }%

\author{Frederico P. Delgado}
\affiliation{%
 Physics Department, TUM School of Natural Sciences, Technical University of Munich, 85748 Garching, Germany
 }

\author{Martin Schwade}
\affiliation{%
 Physics Department, TUM School of Natural Sciences, Technical University of Munich, 85748 Garching, Germany
 }

\author{Peter Beck}
\affiliation{%
 Department of Energy Conversion and Storage, Technical University of Denmark, DK-2800 Kgs.~Lyngby, Denmark
 }%
 
\author{Jonas A. Oldenstaedt}
\affiliation{%
 Physics Department, TUM School of Natural Sciences, Technical University of Munich, 85748 Garching, Germany
 }
 
\author{Ivano E. Castelli}
\email{ivca@dtu.dk}
\affiliation{%
 Department of Energy Conversion and Storage, Technical University of Denmark, DK-2800 Kgs.~Lyngby, Denmark
 }%
 
\author{David A. Egger}
\email{david.egger@tum.de}
\affiliation{%
 Physics Department, TUM School of Natural Sciences, Technical University of Munich, 85748 Garching, Germany
}%
\affiliation{%
 Atomistic Modeling Center, Munich Data Science Institute, Technical University of Munich, 85748 Garching, Germany
 }
\date{\today}% 
             
\begin{abstract}
Predicting optoelectronic response across thermodynamic state space requires coupling finite-temperature nuclear dynamics to electronic structure at scales where direct first-principles calculations are impractical. Machine-learning force fields and Hamiltonian-learning models provide scalable predictions, but integrating them into reliable and interpretable workflows remains challenging. Here, we introduce FLOW-OTTER, a modular, model-agnostic framework that automates molecular dynamics, Hamiltonian prediction, observable extraction, reliability assessment, and Hamiltonian-level interpretation. We demonstrate FLOW-OTTER in halide perovskites, soft semiconductors whose anharmonic fluctuations strongly modulate electronic response. Using FLOW-OTTER, we show that independently learned nuclear and electronic models remain predictive when composed end-to-end, reproducing experimental temperature- and pressure-dependent band-gap trends using models trained only on first-principles targets at zero pressure. By resolving the nonlinear evolution of Pb-$s$/Br-$p$ antibonding at the valence-band maximum, it identifies the microscopic origin of the asymmetric pressure response. FLOW-OTTER thus establishes Hamiltonian learning as a general route from thermodynamic trajectories to experimentally grounded optoelectronic mechanisms.
\end{abstract}


\maketitle

%\tableofcontents

\section*{\label{sec:00_introduction}Introduction}

Atomistic simulations increasingly rely on machine-learning (ML) models to replace expensive first-principles calculations. 
However, predicting experimentally relevant observables often requires multiple learned components to be composed and validated within a single workflow. This challenge is particularly acute in soft semiconductors, where functional properties emerge from ensembles of thermally accessible configurations rather than a single equilibrium structure.
Finite temperature, pressure, disorder, and local structural fluctuations reshape orbital hybridization, band edges, optical response, and charge transport, making optoelectronic prediction a coupled nuclear--electronic problem.
Direct first-principles molecular dynamics (MD) combined with repeated electronic-structure calculations capture these effects in principle \cite{Zacharias2020,seidl2023,Hegner2024}, but their computational cost makes systematic exploration of thermodynamic and compositional state spaces impractical.

ML force fields (MLFFs) have transformed atomistic simulation by enabling MD at length and time scales that are inaccessible to conventional first-principles approaches while retaining near first-principles accuracy within validated regimes \cite{Jinnouchi2022,batatia_mace_2022,Batzner2022}. 
MLFFs provide realistic thermally sampled structures but do not provide the electronic information needed to predict optoelectronic properties on their own. 
Hamiltonian learning has recently emerged as a complementary approach: instead of predicting individual scalar properties, it learns the quantum-mechanical electronic Hamiltonian, a physically structured representation from which optoelectronic observables can be calculated \cite{Li2022,Zhang2022,Zhong2023,Qian2026,Schwade2026}.
Recent developments have broadened the scope of Hamiltonian learning towards more transferable representations, alternative electronic-structure formulations, and Hamiltonian-derived response properties \cite{Wang2024,Gong2024PlaneWave,Zhong2024EPC,Zhong2026}, with the broader development of ML-based electronic-structure methods reviewed in ref.~\cite{Tang2025DeepLearningESC}
Together, MLFFs and Hamiltonian-learning models provide a route to large-scale, finite-temperature simulations that connect atomic motion with electronic structure. 
Realizing their broader scientific value, however, requires more than accurate models in isolation: these learned representations must be composed into reliable, physically interpretable workflows that can systematically explore thermodynamic state space and propagate atomistic dynamics into experimentally relevant electronic observables.

Recent work has begun to bridge atomic and electronic descriptions at the model level. 
For example, MLFFs combined with Wannier-based Hamiltonian prediction can unify structural relaxation and electronic band-structure calculations within a physics-informed model \cite{Qi2025WANDER}. 
Furthermore, coupling equivariant MLFF and Hamiltonian-learning models in an active-learning approach enabled predictions of temperature-dependent electronic properties of defective semiconductors directly from MD trajectories \cite{Zhu2026Defects}.
These advances demonstrate the potential of coupled learned nuclear and electronic models, but they do not yet provide a model-agnostic workflow that combines end-to-end validation of composed models against experimentally accessible observables with Hamiltonian-level mechanistic interpretation.
Together, these gaps define two complementary requirements for coupled ML simulation:
predictive reliability across model boundaries and physically grounded interpretation of the resulting response.

The first challenge is to determine whether independently learned nuclear and electronic descriptions remain reliable when composed into a complete prediction pipeline. Evaluating each surrogate only against its direct computational targets does not establish whether predictive accuracy is retained when thermally sampled structures are mapped onto electronic Hamiltonians and then into emergent observables. A reusable workflow must therefore compose model components reproducibly and assess the coupled pipeline at the level of experimentally measurable response.

The second, and more fundamental, challenge is to translate predictive output into mechanistic understanding.
Learned trajectories and Hamiltonians provide rich electronic information, but they do not by themselves reveal which atomic motions, local distortions, symmetry-breaking events, or orbital rearrangements govern the overall optoelectronic response. 
A useful scientific framework should therefore deliver interpretable, atomistically resolved quantities that expose the structural origin of changes in electronic and optoelectronic behavior. 
Addressing this gap is essential for ML to move beyond property prediction towards the identification of mechanistic structure--property relationships and previously unresolved physical mechanisms.

Here we introduce FLOW-OTTER -- \textbf{F}ramework for \textbf{L}ayered and \textbf{O}rganized \textbf{W}orkflows: \textbf{O}ptoelectronics from \textbf{T}rajectory-based \textbf{T}ime-dependent \textbf{E}lectronic Hamiltonian \textbf{R}outines -- a model-agnostic framework for ML-driven finite-temperature electronic-structure prediction and interpretation. FLOW-OTTER connects trajectory generation, Hamiltonian prediction, observable extraction, reliability assessment, and provenance tracking within a reproducible workflow. Its modular design allows different MLFFs, Hamiltonian-learning architectures, material classes, and electronic observables to be incorporated through common workflow interfaces.

FLOW-OTTER addresses two central challenges in coupled ML simulations. First, it enables the complete MLFF--Hamiltonian-learning pipeline to be evaluated at the level of experimentally accessible optoelectronic response. Rather than validating each surrogate only against its direct computational targets, the workflow tests whether learned nuclear and electronic descriptions retain predictive accuracy when thermally sampled structures are transformed onto electronic Hamiltonians and subsequently into emergent observables that neither model was trained to reproduce. Second, FLOW-OTTER translates predictive output into mechanistic insight by analyzing the learned Hamiltonians with established electronic-structure tools, including projected density of states (PDOS) and crystal orbital Hamilton population (COHP) analysis \cite{Dronskowski1993COHP}. The same learned Hamiltonians used for prediction can therefore be interrogated to reveal how structural fluctuations modify orbital character and bonding interactions, linking changes in atomic structure to shifts in band edges and band gaps.

We demonstrate FLOW-OTTER using halide perovskites, a representative class of soft semiconductors whose anharmonic lattices make their optoelectronic properties highly sensitive to finite-temperature structural fluctuations \cite{Yaffe2017,Marronnier2018Anharmonicity,Gehrmann2019,Schilcher2,Zacharias2023,seidl2023,Frenzel2023,caicedo-davila_disentangling_2024,HyltonFarrington2024,Park2025,Zhu2025,Steele2026}. Using pressure as a controlled perturbation of nuclear structure without changing chemical composition, we predict temperature- and pressure-dependent band-gap trends that agree with experiment, although the underlying Hamiltonian models were trained only on first-principles targets at zero pressure and not on the corresponding band-gap coefficients. Hamiltonian-level analysis further shows that the nonlinear pressure--temperature evolution of Pb-$s$/Br-$p$ antibonding at the valence-band maximum governs the asymmetric band-gap response of CsPbBr$_3$. FLOW-OTTER thus provides a general framework that connects experimentally testable predictions with atomistically resolved mechanisms across the thermodynamic state space.

\section*{\label{sec:01_results}Results}

\subsection*{An automated workflow for coupled nuclear--electronic simulation}
\label{sec:01a_flow_otter}

\begin{figure*}
\includegraphics[width=0.7\textwidth]{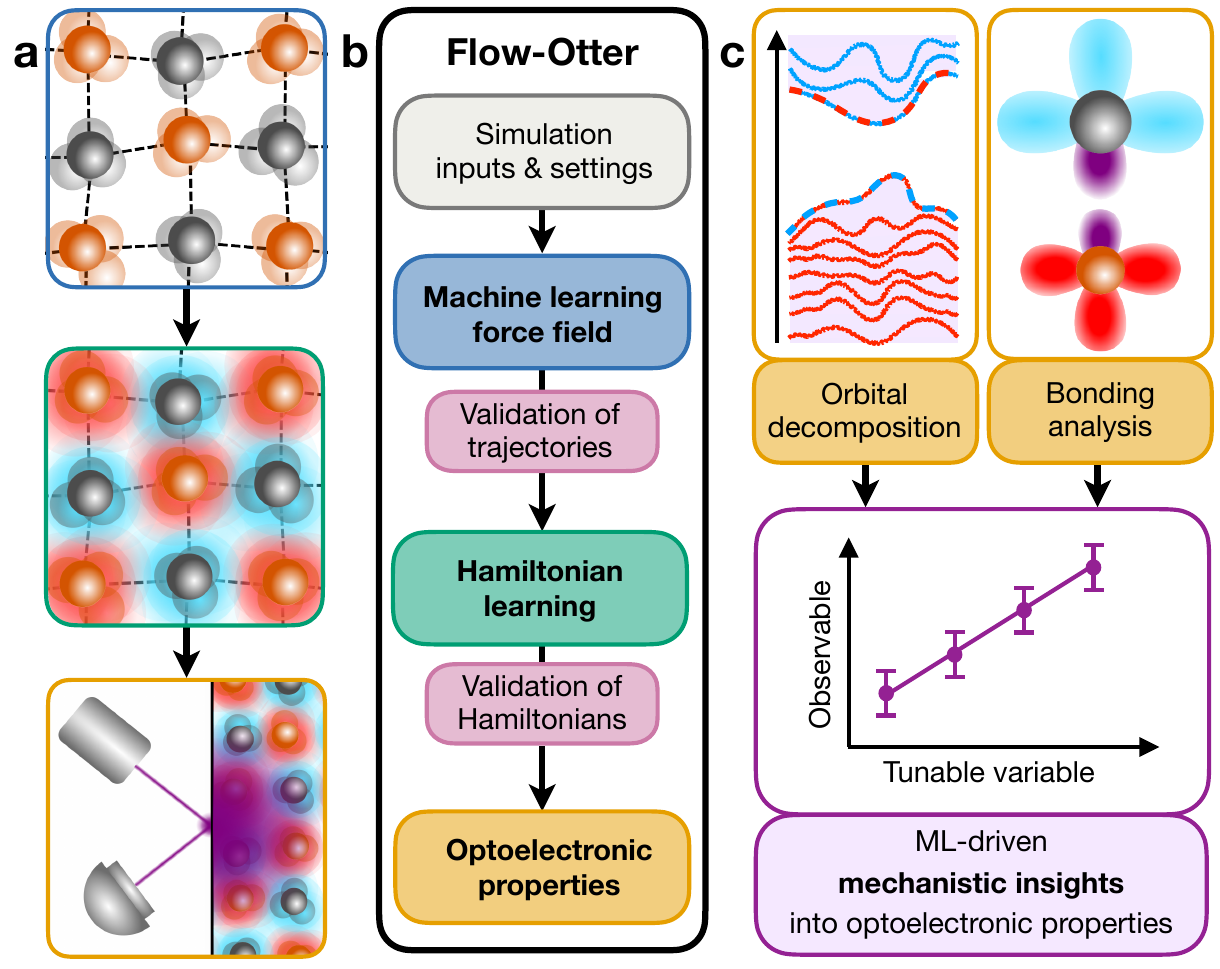} 
\caption{\textbf{Overview of FLOW-OTTER.} \textbf{a} Illustration of the computational challenges associated with realistic finite-temperature prediction of optoelectronic properties. The disordered and fluctuating nuclear structure induces a dynamically evolving electronic structure that governs the material's optoelectronic response. \textbf{b} Schematic flowchart of the automated workflow, FLOW-OTTER, addressing the computational challenges shown in panel a. Based on the simulation setup, machine-learning methods sequentially predict the dynamic nuclear and electronic structure, with each prediction afterward automatically validated by FLOW-OTTER. The resulting confirmed physical description is then used to calculate the material's optoelectronic properties. \textbf{c} Conceptual visualization of FLOW-OTTER's ability to uncover the microscopic origin of changes in measurable optoelectronic properties from interpretable machine-learning outputs by analyzing their orbital and bonding characteristics.}
\label{fig:visualization}
\end{figure*}

We designed FLOW-OTTER as an automated workflow for finite-temperature optoelectronic simulations, in which thermally sampled nuclear configurations are propagated into electronic Hamiltonians, experimentally comparable observables, and mechanistic analyses at the Hamiltonian level. Figure~\ref{fig:visualization}a summarizes the computational challenge addressed by the workflow. Under realistic conditions, finite temperature, pressure, disorder, and local structural fluctuations, one needs to treat an ensemble of atomic configurations rather than a single representative structure. Because optoelectronic response is governed by quantum-mechanical interactions, its prediction requires access to an electronic Hamiltonian, or an equivalent representation of the underlying electronic structure, for each configuration in the ensemble. Evaluating this information directly from first principles across many thermally sampled snapshots is computationally prohibitive. Hamiltonian-learning models address this bottleneck by mapping the atomic structure of each configuration directly onto its corresponding electronic Hamiltonian.

FLOW-OTTER places the learned Hamiltonian at the center of the workflow. Rather than using machine learning to predict only a final scalar quantity, such as a band gap, it maps each sampled nuclear configuration onto a physically structured electronic Hamiltonian that can be propagated into multiple observables and mechanistic analyses. This design turns Hamiltonian learning into a reusable interface between optoelectronic property prediction and microscopic interpretation: the same predicted Hamiltonians can be used to compute band structures, band gaps, densities of states, orbital characters, and bonding descriptors. By retaining this Hamiltonian-level representation throughout the pipeline, FLOW-OTTER enables a coupled nuclear--electronic workflow that is both predictive and mechanistically interpretable.

Figure~\ref{fig:visualization}b shows how FLOW-OTTER implements this approach by coupling learned nuclear dynamics to learned electronic structure in a single automated workflow. Its modular interfaces allow different MLFFs and Hamiltonian-learning models to be incorporated without changing the overall workflow architecture. Starting from structural input and user-defined thermodynamic conditions, the workflow first uses MD to generate thermodynamic ensembles of realistic and reliable atomic configurations.
The resulting nuclear trajectories include effects from thermal motion and disorder or external stimuli such as varying pressure. These configurations are then passed to the Hamiltonian-generation stage, where FLOW-OTTER constructs configuration-dependent electronic Hamiltonians using a chosen Hamiltonian-learning model. In this way, FLOW-OTTER combines the complementary roles of MLFFs and Hamiltonian learning: MLFFs provide scalable access to nuclear configurational ensembles, while Hamiltonian learning converts those configurations into reusable electronic-structure representations. 

The workflow undergoes a sequence of prediction, validation, and interpretation layers. FLOW-OTTER first evaluates the reliability of MD trajectories before generating electronic Hamiltonians, thereby avoiding the computational cost of Hamiltonian prediction from insufficiently equilibrated or poorly sampled trajectories. 
Once a trajectory passes these automated diagnostics, FLOW-OTTER generates configuration-dependent Hamiltonians and assesses their physical plausibility before downstream analysis.
Validated Hamiltonians are then passed to property-prediction modules that compute experimentally comparable observables, including band gaps and densities of states. 
In parallel, interpretation modules extract electronic-structure descriptors such as PDOSs and COHPs. 
These analyses relate predicted optoelectronic observables to orbital character and bonding interactions, establishing crucial mechanistic links between local structural distortions and the resulting electronic response.

In the present implementation, the central Hamiltonian-learning component is HAMSTER, a physics-informed framework that combines an approximate physically motivated Hamiltonian with learned environment-dependent corrections \cite{Schwade2026}. Predicting corrections to a physical baseline rather than the Hamiltonian entirely from scratch, HAMSTER preserves a connection between learned matrix elements and chemically interpretable quantities such as onsite terms, hopping interactions, orbital characters, and bonding environments. This physics-informed Hamiltonian representation is central to FLOW-OTTER because it integrates prediction and interpretation into a single workflow: the predicted Hamiltonians yield physical observables that can be compared to experiments while retaining the information needed to explain their microscopic origin.

The complete workflow is configured via a single input specification that defines structural inputs, thermodynamic conditions, simulation parameters, model choices, analysis settings, data-management options, and parameter sweeps. FLOW-OTTER is implemented using the modular workflow manager PerQueue \cite{perqueue}, which enables persistent workflows with restartable execution, provenance tracking, and dynamic coordination of sequential dependencies between MD, Hamiltonian generation, property extraction, validation, and analysis. Dynamic task groups in PerQueue allow FLOW-OTTER to execute parallel parameter sweeps over user-defined inputs, enabling systematic comparisons across materials, thermodynamic states, computational methods, hyperparameters, and physical parameters.

Figure~\ref{fig:visualization}c illustrates how FLOW-OTTER turns coupled MLFF--Hamiltonian predictions into experimentally testable optoelectronic observables. Rather than validating each learned surrogate only against computational targets, the workflow tests whether predictive accuracy is retained as nuclear configurations are transformed into Hamiltonian prediction and, subsequently, into emergent observables that can be compared directly with experiment. After the internal validation steps described above, the workflow computes optoelectronic observables such as pressure- or temperature-dependent band gaps and densities of states, enabling system-level validation of the complete simulation pipeline. Optional manual inspection points can be inserted before the workflow proceeds, allowing high-throughput automation to be combined with user-controlled reliability checks in applications requiring increased scrutiny.

Beyond prediction, FLOW-OTTER provides a complementary Hamiltonian-level interpretation layer because it is built on a physics-informed Hamiltonian learning model. By retaining Hamiltonian-level information throughout the workflow, it enables predicted agreement or disagreement with experiment to be analyzed in terms of orbital character, band-edge shifts, onsite and hopping terms, and bonding interactions.
The predicted Hamiltonians therefore serve both as generators of measurable observables and as physically structured representations from which microscopic mechanisms can be extracted. For example, by interpreting the Hamiltonians, finite-temperature structural fluctuations can be connected directly to changes in orbital hybridization or bonding. This prediction-and-interpretation strategy is independent of the specific MLFF, Hamiltonian-learning architecture, material system, or observable, provided that the required workflow interfaces are available. FLOW-OTTER thus connects experimentally testable predictions with atomistic explanations, providing a general computational framework for ML-driven scientific discovery rather than a black-box property predictor.

\subsection*{Predicting and validating optoelectronic response across thermodynamic state space}
\label{sec:01b_pressure}

\begin{figure}
\includegraphics[width=0.8\linewidth]{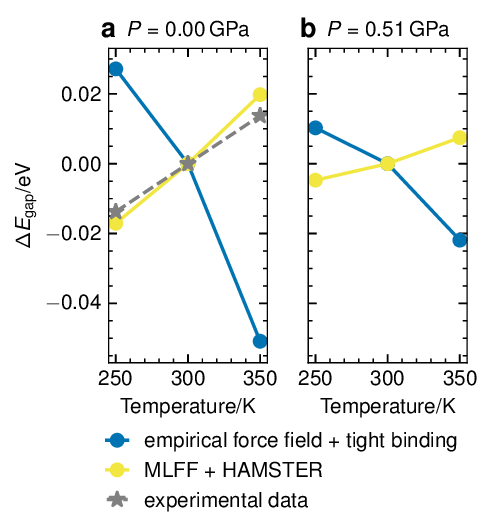}
\caption{\textbf{Validation of temperature-dependent band-gap predictions for MAPbBr$_3$.} \textbf{a} Band-gap change as a function of temperature at \SI{0}{\giga\pascal} for empirical and fully machine-learned model combinations. Grey stars show experimental measurements from ref.~\cite{Mannino2020TemperatureDependent}. \textbf{b} Corresponding predictions at \SI{0.51}{\giga\pascal}.
Data are averaged over 20 configurations from the corresponding trajectory.}
\label{fig:validation}
\end{figure}

Figure~\ref{fig:validation} benchmarks different combinations of MD and electronic-Hamiltonian models within FLOW-OTTER using halide perovskites as a challenging testbed owing to their soft lattices and dynamic disorder.
By exchanging the nuclear and electronic models independently within the same pipeline, this comparison directly demonstrates the modularity of the workflow, with additional model combinations shown in Supplementary Fig.~1.
Selected combinations are then compared against experimental measurements of the temperature-dependent band gap, an emergent observable that neither model was trained to reproduce.
Specifically, we investigate temperature-dependent band gaps of the material MAPbBr$_3$ from MD trajectories at pressures of \SI{0}{\giga\pascal} and \SI{0.51}{\giga\pascal}. We evaluate combinations based on a fine-tuned MACE force field \cite{Batatia2025} and alternative MD models, coupled to HAMSTER \cite{Schwade2026} or semi-empirical electronic Hamiltonians, with additional model combinations reported in the Supplementary Material (SM).

At \SI{0}{\giga\pascal}, Fig.~\ref{fig:validation}a compares the predicted thermal band-gap evolution with experiment \cite{Mannino2020TemperatureDependent}. The combination of the MAPbBr$_3$-specific classical force field \cite{hata_development_2017} and the tight-binding Hamiltonian of refs.~\cite{Mayers,Schilcher2,Vonhoff2025} predicts a negative temperature coefficient, in contrast to the experimentally observed moderate increase of the band gap with increasing temperature. By contrast, the MACE--HAMSTER combination closely reproduces the measured thermal evolution of the band gap, including its positive temperature coefficient. 

This comparison provides a system-level test of the coupled workflow. Neither MACE nor HAMSTER was trained to reproduce the temperature coefficient of the band gap. 
Rather, the observed trend emerges only after thermally sampled structures are mapped onto electronic Hamiltonians and subsequently converted into band-gap predictions. Agreement with experiment therefore assesses whether independently learned nuclear and electronic representations remain compatible when composed into a prediction pipeline, rather than validating either model only against its direct computational targets.

Figure~\ref{fig:validation}b examines whether the contrasting thermal trends of the model combinations persist under compression. At \SI{0.51}{\giga\pascal}, the classical force-field--tight-binding combination continues to predict a negative temperature coefficient, whereas MACE--HAMSTER retains the positive trend obtained at \SI{0}{\giga\pascal}. Because temperature-dependent experimental data are unavailable at this pressure, this comparison does not constitute an independent experimental validation. Instead, it provides a consistency test across thermodynamic conditions, showing that the model ranking established against experiment at \SI{0}{\giga\pascal} is preserved under compression. Therefore, these benchmarks establish confidence in using MACE--HAMSTER for the following exploration of pressure--temperature space and for the Hamiltonian-level analysis of the resulting band-gap response.

\begin{figure}
\includegraphics[width=0.8\linewidth]{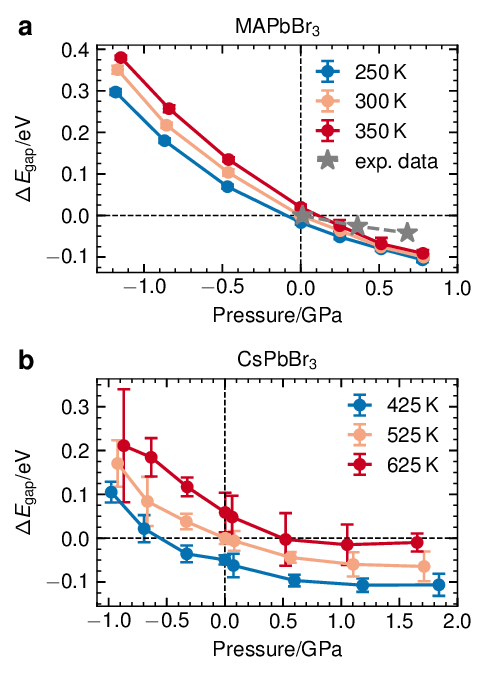} 
\caption{\textbf{Band-gap response across pressure-temperature space in cubic halide perovskites.} \textbf{a} Relative band-gap change of cubic MAPbBr$_3$ as a function of tensile and compressive pressure at three temperatures. Grey stars show experimental measurements at \SI{300}{\kelvin} from ref.~\cite{Wang2015PressureInduced}. \textbf{b} Corresponding pressure--temperature response of cubic CsPbBr$_3$. Data are averaged over 20 configurations from the corresponding trajectory.}
\label{fig:P_vs_gap}
\end{figure}

Figure~\ref{fig:P_vs_gap} demonstrates how FLOW-OTTER extends learned nuclear and electronic representations beyond their ambient-pressure training conditions to predict optoelectronic response across pressure--temperature space. We consider cubic MAPbBr$_3$ and CsPbBr$_3$ under tensile and compressive pressures at three temperatures, restricting the analysis to thermodynamic conditions under which the cubic phases remain stable.

For MAPbBr$_3$, Fig.~\ref{fig:P_vs_gap}a shows that the band gap decreases with increasing pressure and with decreasing temperature, despite both the nuclear and electronic models being trained only at ambient pressure. The workflow therefore reproduces the experimentally established signs of the pressure and temperature coefficients \cite{Wang2015PressureInduced,szafranski_photovoltaic_2017,Mannino2020TemperatureDependent,Celeste2022} across thermodynamic conditions not directly represented in the training data. 
Alternative combinations of nuclear and electronic models show larger deviations from the experimental pressure dependence of the band gap (Supplementary Fig.~2), providing an additional system-level assessment of the coupled pipeline and demonstrating that predictive accuracy depends on the compatibility of the composed models. 
The variation is dominated by shifts in the valence-band maximum (VBM), whereas the conduction-band minimum (CBM) is comparatively less sensitive to pressure (see SM). This response is consistent with the strongly antibonding character of the VBM \cite{Meloni2016,Saleh2021Methylammonium}: increased orbital overlap at higher pressure or lower temperature raises the energy of these states and narrows the band gap \cite{Huang2018}. FLOW-OTTER also predicts an asymmetric response, with a larger band-gap change under tensile than under compressive pressure, consistent with experimental observations for MAPbI$_3$ \cite{Chakrabarti2024}. The separation between the temperature-dependent curves likewise increases in the tensile regime.

Figure~\ref{fig:P_vs_gap}b shows that this predictive capability transfers to the alternative perovskite material CsPbBr$_3$. FLOW-OTTER recovers a positive temperature coefficient and a negative pressure coefficient over tensile and weakly compressive pressures, in agreement with the experimentally reported signs of both trends \cite{Mannino2020TemperatureDependent,Gong2022,Zhang2024}. It also reproduces two material-specific features: a weaker tensile-pressure response than in MAPbBr$_3$ and the onset of band-gap saturation above approximately \SI{1.0}{\giga\pascal}, as observed experimentally \cite{Gong2022,Zhang2024}. These results show that the coupled workflow captures both shared and composition-dependent optoelectronic responses across thermodynamic state space.

The saturation in CsPbBr$_3$ is consistent with two competing structural effects previously identified in orthorhombic CsPbBr$_3$ \cite{Chen2024}. Compression shortens bond lengths and enhances antibonding orbital overlap, which lowers the band gap. At the same time, increased octahedral tilting displaces the halide atoms away from the Pb--Pb bond axis, reduces the alignment of Br-$p$ orbitals and counteracts the increase in antibonding overlap. Because octahedral tilting is more pronounced in CsPbBr$_3$ \cite{Mannino2020TemperatureDependent}, these effects partially cancel, consistent with the saturation in Fig.~\ref{fig:P_vs_gap}b, whereas the band gap of MAPbBr$_3$ continues to decrease. However, the band-gap trends alone do not resolve the microscopic origin of the pronounced asymmetry between tensile and compressive pressure. We therefore interrogate the learned Hamiltonians directly through COHP analysis.

\subsection*{Hamiltonian-level origins of pressure-dependent band-gap response}
\label{sec:01c_bonding}

\begin{figure}
\includegraphics[width=0.8\linewidth]{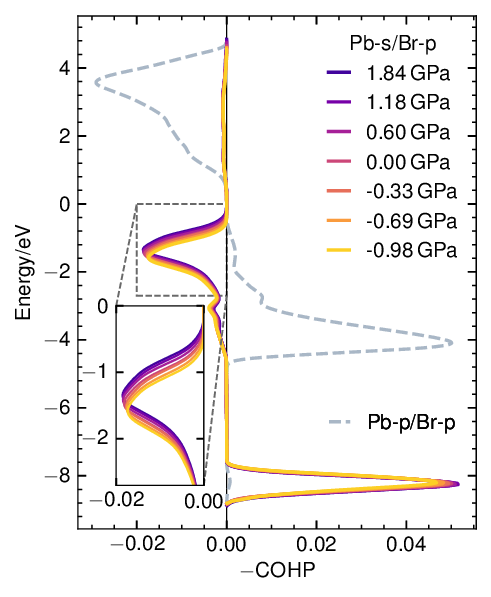} 
\caption{\textbf{Pressure-dependent Pb-Br bonding interactions in CsPbBr$_3$ at \SI{425}{\kelvin}.} Crystal orbital Hamilton population curves are shown for the pressures considered in Fig.~\ref{fig:P_vs_gap}b, with the Fermi level located near \SI{0}{\electronvolt}. Solid lines show Pb-$s$/Br-$p$ interactions, and the dashed line shows the Pb-$p$/Br-$p$ interaction at \SI{0}{\giga\pascal}.}
\label{fig:cohp}
\end{figure}

Figure~\ref{fig:cohp} investigates the Hamiltonian-level origin of the pressure-dependent band-gap response in CsPbBr$_3$. We analyze the Pb-$s$/Br-$p$ interaction across the same pressure range considered in Fig.~\ref{fig:P_vs_gap} and averaged over 20 configurations from the corresponding trajectory, with the Pb-$p$/Br-$p$ interaction included for comparison. The COHPs are normalized by the number of contributing bonds and plotted as negative COHP, so that antibonding contributions appear at negative values (see Methods section for details).

Guided by the established band-edge orbital character \cite{Brenner2016} and our PDOS analysis (see SM), we focus on Pb-$s$/Br-$p$ interactions at the VBM and Pb-$p$/Br-$p$ interactions at the CBM. 
Previous work has connected temperature- and pressure-induced structural changes in halide perovskites to antibonding band-edge states \cite{Meloni2016,Mannino2020TemperatureDependent,Saleh2021Methylammonium,Chen2024}. 
Here, we resolve their joint evolution across finite-temperature tensile and compressive regimes and quantitatively relate it to the asymmetric band-gap response. The corresponding COHPs show that the CBM is associated primarily with antibonding Pb-$p$/Br-$p$ interactions, while the VBM is governed by antibonding Pb-$s$/Br-$p$ interactions. Interestingly, we found that accurately resolving this contribution requires the deep Pb-$s$ states to be included in the HAMSTER training data, and show in the SM that their omission prevents the relevant bonding response from being captured correctly.

\begin{figure}
\includegraphics[width=0.8\linewidth]{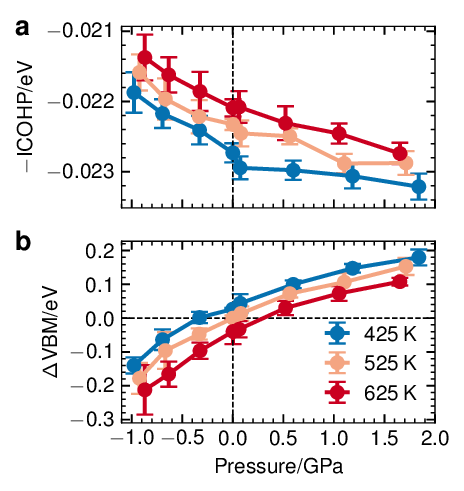} 
\caption{\textbf{Nonlinear Pb-$s$/Br-$p$ antibonding underlies the asymmetric pressure response of the VBM in CsPbBr$_3$.} \textbf{a} Pressure dependence of the Pb-$s$/Br-$p$ COHP, integrated from $-7$ to \SI{0}{\electronvolt}, and plotted as $-$ICOHP. \textbf{b} Corresponding pressure-induced shift of the valence band maximum relative to its zero-pressure value at \SI{525}{\kelvin}. Data are averaged over 20 configurations from the corresponding trajectory.}
\label{fig:icohp}
\end{figure}

Because the pressure dependence of the band gap is dominated by the VBM (see SM), we focus on the Pb-$s$/Br-$p$ interaction. The inset of Fig.~\ref{fig:cohp} highlights the antibonding feature closest to the VBM. With increasing pressure, this feature broadens, increases in magnitude, and shifts towards higher energies, demonstrating that compression strengthens the antibonding interaction governing the valence-band edge. By comparison, the Pb-$p$/Br-$p$ contribution associated with the CBM changes less strongly with pressure.

Figure~\ref{fig:icohp}a quantifies this bonding response through the pressure dependence of the integrated $-$COHP over the antibonding valence-band region.
Its negative values confirm the antibonding character of the VBM, while the increase in absolute magnitude with pressure and its decrease with temperature show that compression strengthens, and thermal expansion weakens, this interaction. Importantly, its pressure dependence becomes less pronounced in the compressive regime, particularly at lower temperatures.

The same nonlinear response is observed in the VBM shifts shown in Fig.~\ref{fig:icohp}b. Relative to their zero-pressure values, the VBM energies rise with pressure as Pb-$s$/Br-$p$ antibonding strengthens. The reduced rate of VBM increase at positive pressure follows the corresponding evolution of the integrated COHP, accounting for the weaker band-gap response under compression than under tension in Fig.~\ref{fig:P_vs_gap}. FLOW-OTTER thus traces the asymmetric finite-temperature band-gap response across tensile and compressive regimes to the nonlinear pressure dependence of Pb-$s$/Br-$p$ antibonding at the VBM.

\hspace{1em}
\section*{\label{sec:02_discussion}Discussion}

We have introduced FLOW-OTTER, a modular framework that connects learned nuclear dynamics, electronic Hamiltonians, experimentally accessible observables, and Hamiltonian-level interpretation within a single automated workflow. In halide perovskites, FLOW-OTTER reproduces temperature- and pressure-dependent band-gap trends, predicts quantities beyond the constituent model training targets, and traces the asymmetric pressure response of CsPbBr$_3$ to the nonlinear evolution of Pb-$s$/Br-$p$ antibonding at the valence-band maximum.

Previous experimental and theoretical studies have related the unusual pressure- and temperature-dependent band gaps of halide perovskites to antibonding band-edge states, bond-length contraction, and octahedral tilting \cite{Meloni2016,Huang2018,Chen2024,Mannino2020TemperatureDependent,Celeste2022}. FLOW-OTTER extends this picture by resolving the finite-temperature evolution of these interactions across tensile and compressive pressure ranges and quantitatively linking their nonlinear response to the observed band-gap asymmetry.

The framework remains limited by the accuracy and compatibility of the coupled MLFF and Hamiltonian-learning models and the quality of configurational sampling. 
Its modular interfaces nevertheless allow alternative models, materials, and Hamiltonian-derived observables to be incorporated without changing the overall workflow architecture. Future developments could integrate uncertainty-aware model selection, adaptive reference-data generation, additional modules for optoelectronic property prediction, and on-the-fly model training and refinement to extend reliable predictions across broader thermodynamic and compositional state spaces. FLOW-OTTER thus provides a route from scalable finite-temperature simulation to experimentally grounded microscopic insight.

\section*{\label{sec:03_methods}Methods}

\subsection*{Molecular dynamics}
\label{sec:03a_MD}

\subsubsection*{Trajectory generation and validation}
\label{sec:03a1_MD}

FLOW-OTTER uses LAMMPS \cite{thompson_lammps_2022} as its molecular-dynamics engine while providing a common interface to different force-field models. Currently implemented backends include empirical FFs, VASP-ML \cite{Jinnouchi2022}, and MACE \cite{batatia_mace_2022}.
Users may choose between NpT and NVT equilibration protocols and configure the corresponding LAMMPS settings through the workflow input. Multiple equilibration stages can be combined as required by the simulation protocol.
FLOW-OTTER assesses trajectory equilibration by comparing the average temperature and pressure with their target values and the temperature variance with the expected statistical fluctuations of the sampled ensemble. Furthermore, the workflow automatically computes trajectory diagnostics and structural observables, including the vibrational density of states, mean-squared displacement, radial distribution function, and coordinate-resolved distributions of atomic displacements, velocities, and forces. These quantities provide additional checks of trajectory quality while enabling physical characterization of the sampled nuclear dynamics.

\subsubsection*{MACE force-field training for halide perovskites}
\label{sec:03a2_MD}

For the nuclear-dynamics simulations of CsPbBr$_3$ and MAPbBr$_3$, we developed separate MLFFs based on the MACE architecture \cite{batatia_mace_2022}.

\textit{MAPbBr$_3$.}
To generate the training data, we performed PBE-level DFT calculations with VASP for 20 snapshots of $4\times4\times4$ supercells extracted from empirical-force-field trajectories at \SI{250}{K}, \SI{300}{K}, and \SI{350}{K}.
Using these reference data, we trained a MACE model \cite{batatia_mace_2022} with two message-passing layers, 128 channels, a maximum equivariance of $L=2$, a radial cutoff of \SI{6}{\angstrom}, correlation order 3, and spherical harmonics up to $l=3$.
The model was trained for 1200 epochs with a batch size of 2, using 10\% of the data for validation.
A Huber loss was used with energy, force, and stress weights of 1, 100, and 100, respectively.
During the final 25\% of training, the energy weight was increased to 1000 and stochastic weight averaging was enabled.
Reference energies for isolated atoms were obtained from spin-polarized DFT calculations.

\textit{CsPbBr$_3$.}
Training data were generated iteratively, beginning with Bayesian on-the-fly learning as implemented in VASP \cite{Jinnouchi2022}.
A VASP MLFF was trained during a \SI{100}{\pico\second} NVT trajectory at \SI{625}{\kelvin} for a $4\times4\times4$ cubic supercell.
Default VASP-ML settings were used except for \texttt{ML\_RCUT1} = \SI{8}{\angstrom}, \texttt{ML\_RCUT2} = \SI{5}{\angstrom}, \texttt{ML\_SION1} = 0.5, and \texttt{ML\_SION2} = 0.5.
The resulting first-principles reference data were used to train an initial MACE model, which was then employed in an NpT temperature sweep from \SI{425}{\kelvin} to \SI{625}{\kelvin} over \SI{240}{\pico\second}.
This trajectory provided structures spanning the lattice-parameter and volume range required for the subsequent simulations.
A second MACE model was trained exclusively on these structures and used in an additional active-learning cycle based on the same NpT heating protocol.

The final CsPbBr$_3$ MACE model used two message-passing layers, 128 channels, \texttt{max\_L} = 2, correlation order 3, spherical harmonics up to $l=3$, and a radial cutoff of \SI{6}{\angstrom}.
It was trained for 600 epochs with a batch size of 10 and a 10\% validation split.
A Huber loss was used with energy, force, and stress weights of 1, 100, and 100, respectively.
Stochastic weight averaging was enabled during the final 120 epochs, together with an increase of the energy weight to 1000.
Reference energies for isolated atoms were obtained from spin-polarized DFT calculations.

\subsection*{Electronic Hamiltonians}
\label{sec:03b_H}

\subsubsection*{Hamiltonian generation and validation}
\label{sec:03b1_H}

FLOW-OTTER currently supports two approaches for electronic-Hamiltonian generation: an empirical tight-binding model parametrized from DFT calculations for halide perovskites \cite{Mayers,Schilcher2,Vonhoff2025} and the HAMSTER ML framework \cite{Schwade2026}.
Both approaches map thermally disordered nuclear configurations onto non-interacting electronic Hamiltonians through a common workflow interface.
Whereas the empirical tight-binding model is specific to halide perovskites, HAMSTER is designed for broader chemical applicability.
HAMSTER employs a $\Delta$-ML approach in which environment-dependent corrections are applied to an effective Hamiltonian derived from a fitted tight-binding model, allowing local electronic changes induced by structural distortions to be represented with accuracy approaching the underlying first-principles reference within validated regimes.

FLOW-OTTER automatically evaluates the distributions of on-site and hopping matrix elements and can inspect the DOS and candidate band gaps for selected configurations.
The workflow is terminated if Hamiltonian elements exceed predefined physical bounds, preventing unphysical predictions from being propagated to subsequent property calculations.
For applications requiring additional scrutiny, optional inspection points can require manual approval of the automatically generated MD and Hamiltonian diagnostics before the workflow proceeds.

\subsubsection*{HAMSTER parameterization for halide perovskites}
\label{sec:03b2_H}

For CsPbBr$_3$, the HAMSTER model was optimized following the general procedure of ref.~\cite{Schwade2026}, with modifications to the selected band window, training data, and band weights.
The underlying tight-binding parameters were fitted to the unit-cell band structure using 10 valence and 4 conduction bands, including the deep Pb-$s$-derived bands.
The ML and spin--orbit-coupling (SOC) parameters were subsequently optimized using training and validation sets of 12 structures each, randomly sampled from the \SI{625}{K} MD trajectory reported in ref.~\cite{Schwade2026}.

Optimization was performed in two stages.
First, the ML and SOC parameters were fitted simultaneously.
The SOC parameters were then fixed, while the ML parameters were reinitialized and optimized independently.
Bands near the lower and upper edges of the selected energy window were assigned weights of 0.75, whereas the VBM and CBM bands were assigned a weight of 5.
The resulting parameterization therefore differs from that of ref.~\cite{Schwade2026} in the selected band window, the temperature of the structures used for training and validation, and the band weights.

For MAPbBr$_3$, we used the HAMSTER model reported in ref.~\cite{Schwade2026} without modification.

\subsection*{Electronic observables and Hamiltonian analysis}
\label{sec:03c_opto}

\subsubsection*{Hamiltonian-derived observables}
\label{sec:03c1_opto}

FLOW-OTTER uses the generated electronic Hamiltonians to evaluate band gaps, densities of states (DOS), PDOS, and COHPs.
Depending on the Hamiltonian dimension, the workflow automatically selects the numerical evaluation scheme.
For Hamiltonians with dimension below $10^4$, eigenvalues and eigenvectors are obtained by exact diagonalization, allowing direct evaluation of the corresponding electronic observables.
For larger Hamiltonians, FLOW-OTTER employs a kernel-polynomial-method-based stochastic approach for efficient large-scale evaluation of the DOS, PDOS, COHP, and band edges, as described below.

\subsubsection*{Large-scale Hamiltonian evaluation with KPM}
\label{sec:03c1b_opto}

For large Hamiltonians, full diagonalization becomes computationally prohibitive.
FLOW-OTTER therefore employs the kernel polynomial method (KPM) \cite{weise_kernel_2006} together with a stochastic trace approximation to evaluate the DOS, PDOS, and COHP.
In this formulation, the COHP and PDOS are written as
\begin{equation}
    \mathrm{COHP}_{ij}(E)
    =
    \mathrm{Tr}\left(P_i H P_j \delta(E-H)\right),
    \label{eqn:COHP}
\end{equation}
and
\begin{equation}
    \mathrm{PDOS}_i(E)
    =
    \mathrm{Tr}\left(P_i \delta(E-H)\right),
    \label{eqn:PDOS}
\end{equation}
where $P_i = |i\rangle\langle i|$ projects onto orbital state $|i\rangle$.
Replacing $P_i$ by the identity operator yields the total DOS.

The traces are approximated using stochastic vectors, avoiding construction of a complete basis representation.
Within KPM, the Dirac delta distributions are expanded in Chebyshev polynomials, reducing the calculation to sparse matrix--vector operations rather than matrix--matrix operations.
For fixed expansion parameters, this formulation enables efficient scaling to large Hamiltonians.
Orbital-resolved PDOS and COHP are obtained by summing the corresponding orbital or orbital-pair contributions.

For band-gap calculations with large Hamiltonians, FLOW-OTTER first estimates the VBM and CBM positions from the KPM DOS.
These estimates are then used to compute only the eigenvalues closest to the respective band edges, avoiding full diagonalization while retaining an explicit determination of the band gap.

\subsubsection*{Prediction of pressure- and temperature-dependent band gaps}
\label{sec:03c2_opto}

Pressure- and temperature-dependent electronic properties were evaluated from ensembles of configurations sampled from equilibrated MD trajectories.
All calculations used $16\times16\times16$ supercells to reduce finite-size effects and sample the orientational disorder of the MA sublattice \cite{Schwade2026}.
At each target temperature, the zero-pressure reference structure was obtained by NVT equilibration followed by NpT equilibration.
The average lattice constants from the NpT trajectory were then used to construct a cubic reference cell with isotropic lattice vectors.

Tensile and compressive states were generated by isotropically scaling this reference cell.
For negative pressures, the lattice constants were expanded by \SI{1}{\percent}, \SI{2}{\percent}, and \SI{3}{\percent}.
Positive pressures were restricted to the compression range over which the cubic phase remains stable.
For CsPbBr$_3$, lattice contractions of \SI{1}{\percent}, \SI{2}{\percent}, and \SI{3}{\percent} were applied, remaining below the cubic-to-orthorhombic transition pressure of approximately \SI{2.0}{\giga\pascal} at \SI{300}{K} \cite{Zhang2024}.
For MAPbBr$_3$, smaller contractions of \SI{0.5}{\percent}, \SI{1.0}{\percent}, and \SI{1.5}{\percent} were used because the cubic-to-tetragonal transition occurs at approximately \SI{0.8}{\giga\pascal} at room temperature \cite{Celeste2022}.
For each scaled cell, an NVT production trajectory was generated, and the time-averaged virial pressure reported by LAMMPS was used as the effective pressure of the ensemble.

Electronic properties were evaluated from 20 randomly selected configurations from each production trajectory.
For each configuration, the corresponding electronic Hamiltonian was generated with HAMSTER and used to calculate the band gap.
The thermal band gap was obtained by averaging over the 20 configurations.
Because the Hamiltonians were trained against PBE reference calculations \cite{PBE}, we report relative band-gap changes to remove the systematic offset in the absolute PBE band gap.
The same ensemble-averaging procedure was applied to the DOS, PDOS, and COHP.

\section*{Data availability}

The data that support the findings of this study are available in the Zenodo repository at \url{https://doi.org/10.5281/zenodo.22917148} \cite{data_zenodo}.

\section*{Code availability}

The code used in this study is available at (\url{https://github.com/TheoFEM-TUM/FLOW-OTTER}). The calculations were performed using version 1.0.1, which is archived on Zenodo \cite{code_zenodo}.

\begin{acknowledgments}

We thank Philip Schwinghammer for fruitful discussions about the crystal orbital Hamiltonian population.
Funding provided by Germany's Excellence Strategy – EXC 2089/2-390776260, {the Deutsche Forschungsgemeinschaft (DFG) via SPP2196 Priority Program (Project-ID: 424709454)}, and by the Studienstiftung des Deutschen Volkes, are gratefully acknowledged. The authors further acknowledge the Gauss Centre for Supercomputing e.V. for funding this project by providing computing time through the John von Neumann Institute for Computing on the GCS Supercomputer JUWELS at Jülich Supercomputing Centre.

\nocite{DataAvailability}
\end{acknowledgments}

\bibliographystyle{aipnum4-1}
\bibliography{reference}

\end{document}